\documentclass[12pt]{iopart}

\usepackage[utf8]{inputenc}
\usepackage{amsfonts}
\usepackage{amssymb}
\usepackage{graphicx}
\usepackage{xcolor}

\begin{document}
\title{Defect generation and dynamics during quenching in finite size homogeneous ion chains}

\author{J. Pedregosa-Gutierrez$^{1,2,3}$ and M. Mukherjee$^{2,3,4}$}
\address{$^1$ Aix Marseille Univ, CNRS, PIIM, Marseille, France}
\address{$^2$ MajuLab, International Joint Research Unit UMI 3654, CNRS, Université
Côte d’Azur, Sorbonne Université, National University of Singapore, Nanyang Technological University, Singapore}
\address{$^3$ Centre for Quantum Technologies, National University Singapore, Singapore, 117543, Singapore}
\address{$^4$ Department of Physics, National University Singapore, Singapore, 117551, Singapore}

\ead{jofre.pedregosa@univ-amu.fr}

\date{\today}

\begin{abstract}
An equally spaced linear chain of ions provides a test-bed for studying the defect formation during a topological phase transition from a linear to a zig-zag configuration. By using a particular axial potential leading to an homogeneous ion chain, the boundary conditions are not needed, allowing new rich defect dynamics to appear on an homogeneous system. A semi-empirical expression for the critical transition frequency provides an excellent agreement to the numerical results for low ion numbers. The non-adiabatic crossing of the phase transition shows different power-laws for the defect probability density for different quench rates regions. Information regarding defect dynamics is obtained through the measurement of the defect density at different times during the quench. By comparing the defect density and the correlation length dynamics among the different number of trapped ions, the role of the different defect loss mechanism can be deduced. An excellent agreement with the predictions given by the homogeneous Kibble-Zurek model is found on a finite size system of 30 ion system which can be tested in present ion trap experimental set-ups.
\end{abstract}

\pacs{keywords: radio-frequency ion traps; laser cooling; phase-transition; Kibble-Zurek}

\submitto{\NJP}
\maketitle
\section{Introduction}
Phase transition is ubiquitous in nature as well as in the laboratories, from boiling of water to the formation of Bose-Einstein condensate. Symmetry breaking during the second-order phase transitions have been invoked to explain the observed inhomogeneity of mass distribution of the universe through the Kibble-Zurek mechanism (KZM)\cite{zurek_cosmological_1985, kibble_topology_1976}. However, such a mechanism is universal to any phase transition crossing a critical point, making some of the aspects of the dynamics independent of the underlying system.

The original KZM (also known as the Homogeneous KZM) assumes that the system is initially homogeneous (uniform density) and infinite (no edge boundaries). If the system is driven over a second-order phase transition to a ground-state with multifold degeneracy, the system has to choose one among the different possible final states. In an infinite and homogeneous system the whole system crosses to the new phase simultaneously. The speed at which the new phase choice can be communicated is finite, from which follows that there is a finite probability that different causally disconnected regions will choose different final states, leading to defect formation at the boundaries of different choices. Since these defects are structural in nature, they are known as topological defects. 

One of the quantifiable predictions of the KZM is the number of topological defects formed as a function of the rate at which the critical point of a phase transition is crossed, also called the defect probability density. Such speed is controlled experimentally through the time variation of a control parameter like temperature in case of boiling water. Three main difficulties typically appear when investigating the KZM in the laboratory. First, the weak power law scaling of the defect densities requires to experimentally explore a large range of the control parameter around the critical point. Second, the preparation of an homogeneous initial system and finally, the preparation of a large enough system where finite size effects do not play a significant role.

An overview of the theoretical background of the KZM and of the experimental efforts regarding the KZM can be found in~\cite{del_campo_universality_2014}. To our knowledge, the best agreement between the predicted power law and experimentally measured one is achieved using a linear ion chains formed in linear RF traps~\cite{pyka_topological_2013, ulm_observation_2013}. In this experiment, a finite size chain of ions trapped in a harmonic potential and laser cooled to form a linear chain of 1D Coulomb crystal. The chain was driven to a doubly-degenerate configuration of Coulomb crystal, namely zig-zag and zag-zig, based on whether the ions are push up or down with respect to the linear chain equilibrium point. However, such experiments could not explore the homogeneous case as the ion density in a harmonic ion trap is not constant~\cite{kamsap_experimental_2017}. This experiment was faithfully explained by modifying the original homogeneous case to the inhomogeneous KZM~\cite{del_campo_structural_2010} (IKZM) in which the transition is crossed always at the place of highest density first e.i, at the center of the ion chain, leading to a modification of the scaling law.

As the ion trap technology is advancing rapidly, the HKZM could be studied using an uniform distribution of laser cooled ions. One possibility is the generation of ion rings in multipole traps~\cite{champenois_ion_2010}. The HKZM using an ion ring has been studied before numerically in~\cite{nigmatullin_formation_2016}. However, the experimental realisation of ion rings in multipole traps has proven more difficult than initially expected~\cite{pedregosa-gutierrez_symmetry_2018, pedregosa-gutierrez_correcting_2018}, with only one reported ion ring generated in a multipole surface trap~\cite{li_realization_2017}. A ring satisfies both the criteria namely, homogeneity and infinity. 

Another possibility is to use a linear ion chain with uniform inter ion spacing. Such ion crystal configuration has been proposed in the context of ion trap quantum computing~\cite{lin_large-scale_2009}. Here, we explore the possibility to use such a topology, the homogeneous ion chain of finite size, for the study of the defect dynamics under the ambit of HKZM. The KZM in general and the homogenous KZM in particular are becoming more and more relevant in the context of quantum technology. As an example, mesoscopic systems of uniform lattice structure are generated via classical phase transitions in well controlled experiments in order to develop super-lattice by molecular self-assembly for single photon generation~\cite{raino_superfluorescence_2018, li_structural_2020}. However, such systems are often bugged by unwanted defects leading to inefficient photon generations. In order to understand similar complex systems, it is important to design controlled experiments where defect formation during a phase transition can be be better understood and controlled. One of the simplest but highly controllable system is a linear chain of uniformly spaced trapped and laser cooled ions. A structural or topological phase transition can be driven by varying only the ratio of the confinement strength in two dimensions in a well controlled manner, thus providing a toolbox to verify and control defect formation in any second order phase transition due to the universality of the KZM.

Unlike condensed matter systems, where the size often can be considered infinite but provides limited degree of controllability, an ion chain is a well-controlled Coulomb crystal that can be made homogeneous but finite in size. Finite size systems are going to be relevant in the near term quantum technology, thus knowing finite size effects in HKZM with edge boundaries, is likely to play a significant role. The universality of defect formation as a function of the rate at which the critical point of the phase transition is crossed is rather well known from the theoretical side. Such universality has also been tested for finite size systems with periodic boundary conditions~\cite{nigmatullin_formation_2016,puebla_fokker-planck_2017} and re-scaling has been suggested~\cite{nigmatullin_formation_2016}. However, such studies do not include the possibility of defect loss which can be significant, particularly for the type of small systems which are experimentally accessible for near term quantum technology implementation. Moreover, such losses may even alter the universality of the scaling when experiments are performed to test these universality classes. In particular, we have used molecular dynamics (MD) simulations of the 1D to 2D (linear to zig-zag) transitions of a homogeneous ion chains to study the behaviour of defect formation as a function of the velocity by which the transition is crossed for a finite size system with edges. To the best of our knowledge, this investigation for the first time provides an insight on the losses, both at the boundaries and through annihilations.

The article is organised as follows: first, the electric potential needed to generate an homogeneous ion chain is discussed and verified through MD simulations. A new derivation of the critical parameter at which the linear chain to zig-zag transition occurs and MD simulations to verify the value of the critical parameter in function of the ion-ion distance and the number of ions are discussed next. Followed by a discussion in terms of defect generations of MD simulations of the linear to zig-zag phase transition over a large range of the rate at which the critical parameter is changed. The results are then analysed in the context of the HKZM, followed by a discussion of the correlation length evolution.

\section{Homogeneous Linear Ion Chain}
In a segmented linear ion trap the radial potential is generated by a radio-frequency field while the axial confinement is achieved by appropriate electrostatic voltages applied to the segmented electrodes along the axis. Without segmentation the axial potential is approximately harmonic in nature. However, the electrostatic potential required to generate an homogeneous ion chain can be obtained by finding a ion position dependent potential that compensates all Coulomb forces among the ions at each ion site~\cite{johanning_isospaced_2016}. The resulting analytical solution is:
\begin{eqnarray} \label{eq:homo_pot}
\phi_N(z) = \frac{Q k_C}{d}[ 2\psi^{(0)}(N_+) - \psi^{(0)}(\tilde{z}_+) - \psi^{(0)}(\tilde{z}_-) ] \\
N_+ = (N-1)/2 ~~;~~ \tilde{z}_{\pm} = N_+ \pm z/d, \nonumber
\end{eqnarray}

where the $\psi^{(n)}(x)$ represents the polygamma function, $d$ is the inter-ion spacing, $N$ is the number of ions, $Q$ is the ion's charge and $k_C$ is the Coulomb constant.

A more recent work~\cite{leung_entangling_2018} uses a different approach, where the ion chain is approximated as a uniform charge density $\rho_0 = Q / d$, leading to an expression for the effective electric field acting on each ion due to the Coulomb interaction with all the other ions on the chain. By integrating such an electric field, a different expression than the eqn.~(\ref{eq:homo_pot}) was derived:
\begin{equation}
\phi_{L}(z) = k_C \rho_0 \ln\left( \frac{L^2}{L^2 - z^2} \right)
\label{eq:homo_pot2}
\end{equation}
where $L$ is the half length of the ion chain. The above expression leads to infinite potential walls as noted by the authors. Moreover, such potential leads to a $5\%$ variation in homogeneity on their studied case (number of ions, $N=50$, and $d=3\mu m$)~\cite{leung_entangling_2018}.

In the following, through the use of MD simulations, it is shown that eqn.~(\ref{eq:homo_pot}) leads to a degree of inhomogeneity much lower than $5\%$. For such reason, eqn.~(\ref{eq:homo_pot}) has been used as axial potential in the rest of this work.

The experimental realisation of an homogeneous ion chain has been attempted in Xie et al.~\cite{xie_creating_2017}. A chain of $21$ ions with an ion-ion distance of $8.5 \pm 0.39~\mu$m was achieved using a surface ion trap with multiple electrodes. While the degree of homogeneity achieved at Xie et al~\cite{xie_creating_2017}, $4.6$\%, it is probably not enough for the KZM experiments, it is indeed reasonable to expect an improvement in the near future.

\section{Molecular Dynamic Simulations}
The problem at hand involves solving the equations of motion of $N$ interacting ions in the presence of the trap potential. The laser cooling is represented through a friction term while the different possible heating mechanisms in a real trap are taken into account through a single thermal bath. Such a problem can be described by the Langevin equation for each ion $i$ evolving as:
\begin{eqnarray}
m \partial_{tt} x_i = Q^2 k_C\sum_{j=1,j\neq i}^N{\frac{x_i - x_{j}}{|\vec{r}_{i}-\vec{r}_{j}|^3}} - m\omega_x^2 x_j- \Gamma\partial_t x_i + \sqrt{2\Gamma k_B T} \theta_{xj} \nonumber\\
m \partial_{tt} y_i = Q^2 k_C\sum_{j=1,j\neq i}^N{\frac{y_i - y_{j}}{|\vec{r}_{i}-\vec{r}_{j}|^3}} - m\omega_y^2 y_j- \Gamma\partial_t y_i + \sqrt{2\Gamma k_B T} \theta_{yj} \nonumber \\
m \partial_{tt} z_i = Q^2 k_C\sum_{j=1,j\neq i}^N{\frac{z_i - z_{j}}{|\vec{r}_{i}-\vec{r}_{j}|^3}} - Q\left|\frac{d V(z)}{d z}\right|_{z_i} - \Gamma\partial_t z_i +\sqrt{2\Gamma k_B T} \theta_{zj},
\end{eqnarray}
where $\vec{r} = (x,y,z)$, $\omega_x$ and $\omega_y$ are the secular frequencies along the transverse directions, $x$ and $y$ respectively, $V(z) = \phi_N(z)$ is the axial potential given by~eqn.(\ref{eq:homo_pot}), $\Gamma$ is the friction coefficient, $k_B$ is Boltzmann's constant, $T$ is the temperature, which has been assumed to be the same on all three spatial dimensions and $\theta_{xj}$, $\theta_{yj}$ and $\theta_{zj}$ are a collection of independent standard Wiener processes~\cite{skeel_impulse_2002}. The equations of motion are numerically solved using the vGB82 algorithm as described in~\cite{skeel_impulse_2002}.

MD simulations are used to test the degree of homogeneity of the ion-ion distance on the final ion distribution when the potential given by eqn.~(\ref{eq:homo_pot}) is used. The parameters used are: $\omega_{x}/2\pi = 500~$kHz, $\Gamma = 1.5\cdot10^{-20}$ kg/s and $N=128$ ions of $^{138}$Ba$^+$. The temperature is set to the typical Doppler limit value of $T=0.5$ mK. The ions are initialised as a perfectly homogeneous chain with inter-ion distance $d=10~\mu$m, followed by uniformly distributed random displacement between $\pm1\mu$m is applied in both transverse and axial directions. The ions are evolved during $1$ms with a time step of $2~$ns. If the average position of each ion for the last $0.5~$ms of evolution is taken as their mean position, the final average ion-ion distance is $\langle \Delta z \rangle = 10.000 \pm 0.002~\mu$m, (one standard deviation is used as error measurement) which represent an excellent agreement with the designed inter-ion distance of $10~\mu$m. If the temperature is increased to $5~$mK, the homogeneity remains high with an average ion distance of $\langle \Delta z \rangle = 9.999 \pm 0.003~\mu$m.

\section{Linear to zig-zag transition}
The transition from linear to zig-zag occurs when the control parameter (in this case the radial trap frequency) is modified such that the system crosses the 1D to the 2D topological phase. In the following, we obtain both analytically and numerically, the value of the control parameter at which such a phase transition occurs.

\begin{figure}
 \centering
 \includegraphics[width=1.00\textwidth]{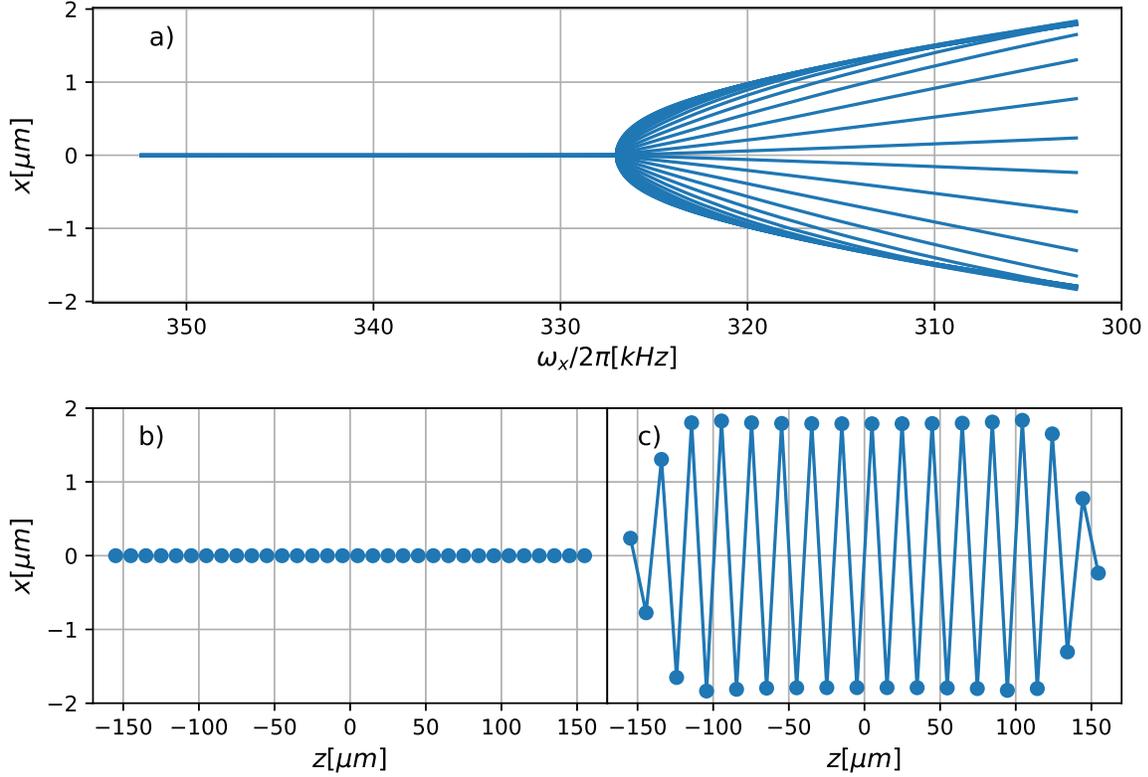}
 \caption{Molecular Dynamic simulation of a quench for a system of 32 $^{138}$Ba$^+$. a) Radial position of each ion vs the $\omega_x(t)$ evolution. b) Ion configuration at $\omega_x(t=0) = 352kHz$. c) Ion configuration at end of the simulation: $\omega_x(t=\tau_Q) = 302kHz$ }
 \label{fig:example_transition}
\end{figure}

In the present system, it is the value of $\omega_x$ that is reduced over time while keeping $\omega_y$ fix, leading to a decrease in the transverse confinement which eventually generates the topological phase jump to a zig-zag structure along the $x$ direction. The value of $\omega_y/2\pi = 1~$MHz is kept constant through the rest of the present work. Such value has been chosen so that $\omega_y > \omega_x$ in the range of $\omega_{x}$ used. Therefore, the transition always occurs along the $x$ direction. An example of such transition for 32 $^{138}$Ba$^+$ ions is shown on figure~\ref{fig:example_transition}, where the transition has been crossed adiabatically. The phase transition is clearly visible in the figure~\ref{fig:example_transition}a, where the radial position as a function of $\omega_x(t)$ for all the 32 ions is plotted. The initial, figure~\ref{fig:example_transition}b), and final positions, figure~\ref{fig:example_transition}c), are also shown. In this example, the simulation parameters are: $d=10\mu m$, $\Gamma = 1.5\cdot10^{-20}$ kg/s and $T=1~$nK.

If periodic boundary conditions are used, the critical frequency, $\omega_c$, at which the transition occurs can be obtained analytically\cite{fishman_structural_2008}:
\begin{eqnarray}\label{eq:wc_official}
\omega_c^2 = \frac{Q^2 k_c}{m d^3} 4\sum_{j=1}^{N}{\frac{1}{j^3} \sin^2{\frac{j\pi}{2}}} \label{eq:wc_official_a} \\
\omega_c^2 (N\to\infty) = \frac{Q^2 k_c}{m d^3} \frac{7\gamma(3)}{2} \label{eq:wc_official_b} \end{eqnarray}
where $\gamma(p)$ is the Riemann-zeta function.

\begin{figure}
 \centering
 \includegraphics[width=1.0\textwidth]{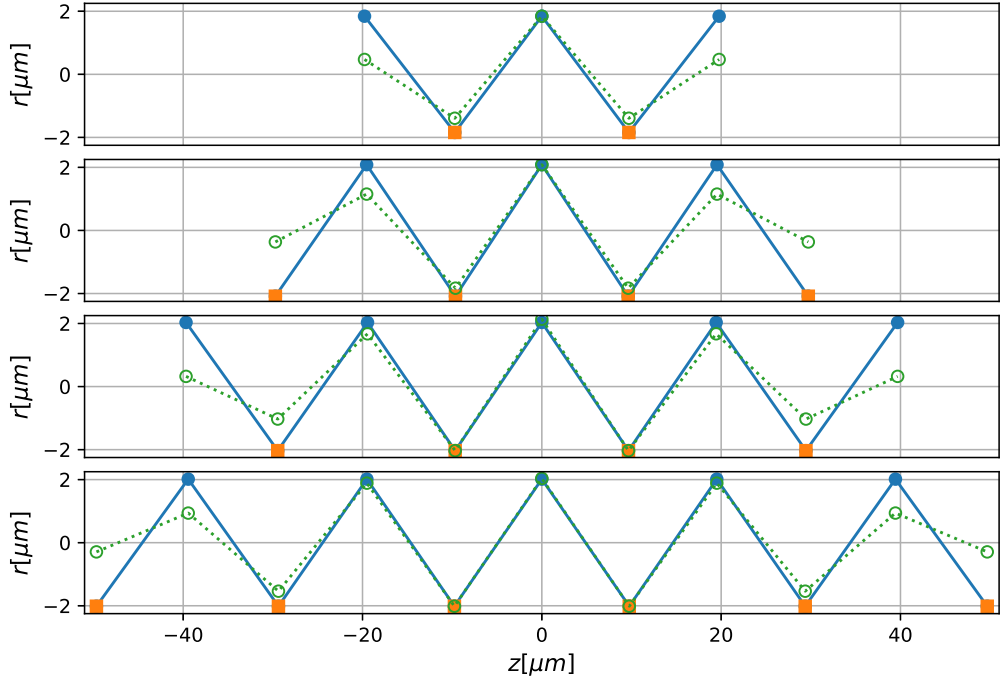}
 \caption{Full Symbols: ion configuration assumed for the derivation of the eq~\ref{eq:wc_odd}. Orange squares: ions with a net contribution on the total force on the central ion. Green empty circles: final ion configuration after a MD simulation where the transition chain to zig-zag is crossed adiabatically}
 \label{fig:ideal_zig_zag2}
\end{figure}

The periodic boundary condition do not correctly represent a true finite size system as defect number remains conserved. In the following, a more realistic situation has been considered, which leads to a better agreement with numerical results as shown in the following.
Let's assume that we have a homogeneous zig-zag ion configuration, as given by the full symbols in figure~\ref{fig:ideal_zig_zag2}.

We assume an odd number of $N$ ions aligned in perfect zig-zag configuration along the $x$ plane ($y=0$), as shown by full circles in figure~\ref{fig:ideal_zig_zag2} with a axial inter-ion distance, $z_{j+1} - z_{j} = d$ and with $|x_j| = h$. By symmetry considerations, only the ions marked with a full square contribute to a net transverse force on the central ion as the axial force contribution of all the other ions cancels out. Therefore, the total force on the central ion can be written as:
\begin{equation}
|\vec{F_{c}}| = 2 Q^2 k_c \sum_{j=1}^{N'}{\frac{2h}{(4h^2 + (2j-1)^2 d^2)^{3/2}}},
\end{equation}
where $N'=|\frac{N+1}{4}|$.

The force due to the confining transverse potential is given by $ |\vec{F_{trap}}| = -m \omega^2 h $. By imposing equilibrium of forces we thus obtain:
\begin{equation}
\omega^2 = 4 \frac{Q^2 k_c}{m} \sum_{j=1}^{N'}{\frac{1}{(4h^2 + (2j-1)^2 d^2)^{3/2}}}
\end{equation}

Taking the limit $\lim h \to 0$, we arrive at:
\begin{equation} \label{eq:wc_odd}
\omega_{c}^2 (N,d) = \frac{Q^2 k_c}{m d^{3}} 4\sum_{j=1}^{N'}{\frac{1}{(2j-1)^{3}}}.
\end{equation}

Notice that eqn.~\ref{eq:wc_odd} leads to a non continuous increase of $\omega_c$ with the number of ions $N$. For example, on figure~\ref{fig:ideal_zig_zag2}, we see that, in the idealised final configuration, the net force on the central ion is exactly the same for the case $N=7$ and $9$ ions.

The MD simulations have been performed in order to check the validity of the above expressions, eqn.~(\ref{eq:wc_official}) and eqn.~(\ref{eq:wc_odd}). The ions are initialised in the chain phase, $\omega_a=\omega_x(t=0) > \omega_c$, with zero velocity. They are first thermalised at the initial radial frequency during $2.0ms$. Then the $\omega_{x}$ frequency is lowered linearly to a final frequency in the zig-zag phase, $\omega_b=\omega_x(t=\tau_Q) < \omega_{c}$ at a speed given by the quench rate: $\frac{\Delta \omega}{\tau_Q}$, where $\Delta \omega = \omega_b - \omega_a$ and $\tau_Q$ is the quench duration:
\begin{equation}
\omega_x(t) =
\left\{
 \begin{array}{ll}
   \omega_a    & \mbox{if } t \leq 0 \\
   \omega_a + \frac{\Delta \omega}{\tau_Q} t & \mbox{if } 0 < t < \tau_Q \\
   \omega_b    & \mbox{if } t \geq \tau_Q .\\
 \end{array}
\right.
\end{equation}

The following MD simulations have been performed for $N=30$ ions and $d=10\mu$m, using a thermal bath temperature of $1$nK. While this temperature represents a non-realistic experimental value, it seemed appropriate to use a very low temperature as eqn.~(\ref{eq:wc_official}) and eqn.~(\ref{eq:wc_odd}) are derived assuming zero temperature. The critical frequency, obtained from eqn.~(\ref{eq:wc_odd}) is $\omega_c = 327.38$kHz. The values used for the initial and the final transverse frequencies are $\omega_b / 2\pi = 140$kHz and $\omega_a / 2\pi = 500~$kHz.

\begin{figure}
 \centering
 \includegraphics[width=1.0\textwidth]{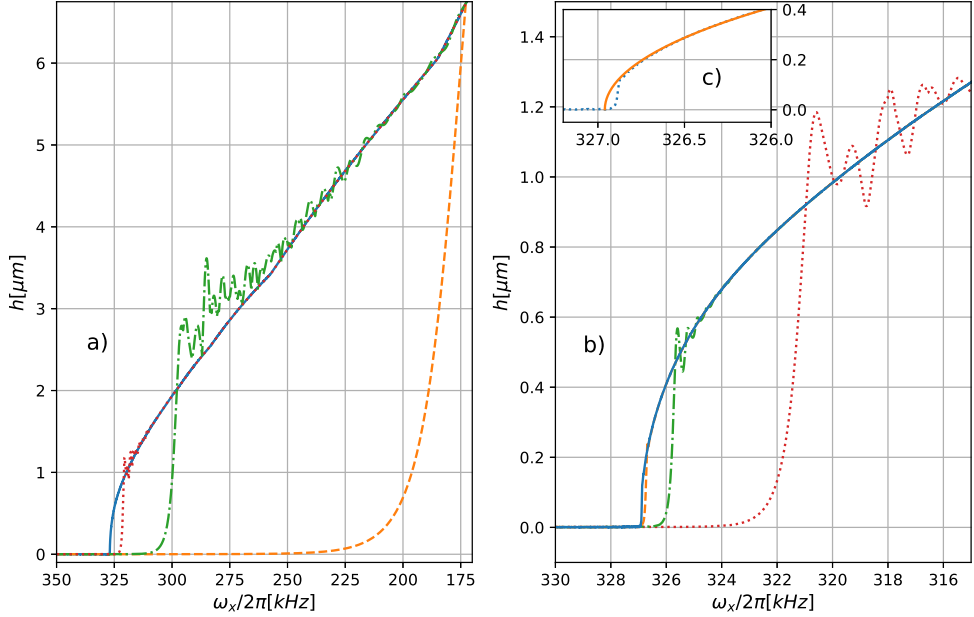}
 \caption{Evolution of the maximum radial extension of the ions configuration ($N=30$, $d=10\mu m$ and $T=1nK$) as a function of the $\omega_x$ for several quench rates, $\frac{\Delta \omega}{\tau_Q}$. a) Dashed orange line: $10^{11}$Hz/s, Dash-dot green line: $10^{10}$Hz/s, Dotted red line: $10^{9}$Hz/s. b) Dotted red line: $10^{9}$Hz/s, Dash-dot green line: $10^{8}$Hz/s, Dashed orange line: $10^{7}$Hz/s, Solid blue line: $10^{6}$Hz/s}
 \label{fig:N32_transition}
\end{figure}

The maximum transverse distance between the two ions $h$, as a function of the instant transverse secular frequency, $\omega_x(t)$, for different values of $\tau_Q$ is shown in figure~(\ref{fig:N32_transition}). Note that the time axis runs from higher values of $\omega_{x}$ to lower ones. Topological phase transitions are clearly observed: for $\omega_x(t)> \omega_{c}$, the value of $h$ is very close to zero (for the slowest quench, the mean value of the $h$ before the transition is $\langle h \rangle = 0.34 \pm 0.17 nm$), which corresponds to a chain configuration, while for $\omega_x(t) < \omega_{c}$ there is a jump on the $h$ value, corresponding to the jump to the zig-zag phase.

Figure~\ref{fig:N32_transition} shows a strong dependency between the exact value of the frequency at which the phase transition occurs, $\omega_{c}$ and the quench rate, $\frac{\Delta \omega}{\tau_Q}$. For the fastest quench studied, $\frac{\Delta \omega}{\tau_Q}=10^{12}$Hz/s, or $\tau_Q=2.26\mu s$, the ions are still in the chain phase at the end of the quench. The observed oscillations following the phase transitions are quickly reduced when the quench time increases. The critical frequency obtained analytically corresponds to the adiabatic case, where $\frac{\Delta \omega}{\tau_Q} \to 0$. The oscillations observed on $h$ just after the transition had been already observed in different system by Shimizu et. al~\cite{shimizu_dynamics_2018, dutta_nonequilibrium_2012} when studying the dynamics of a Mott insulator to a superfluid crossing.

\begin{figure}
 \centering
 \includegraphics[width=1.00\textwidth]{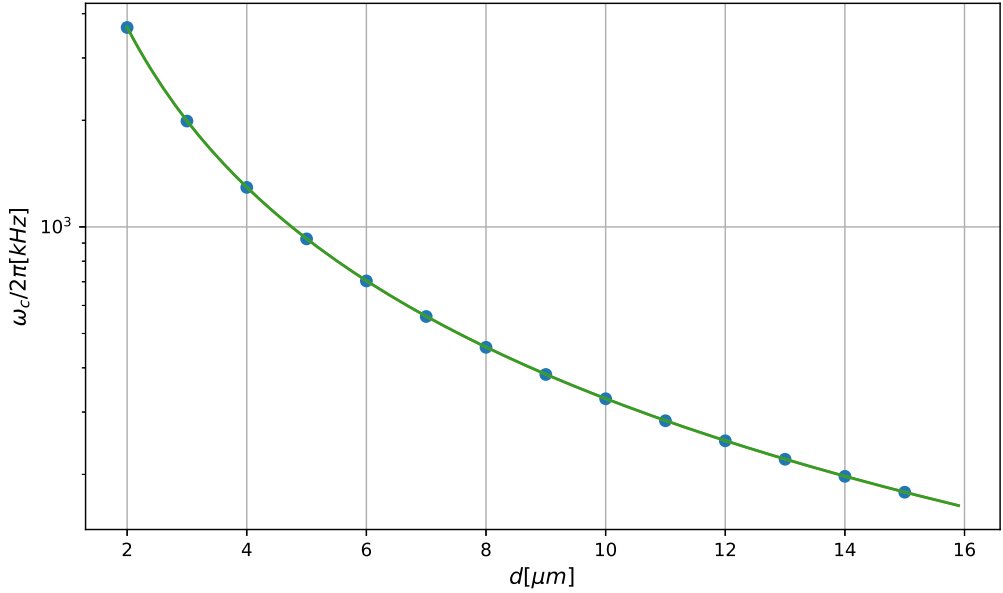}
 \caption{Comparison between the critical frequency obtained from molecular dynamics (circles) and the theoretical value obtained from equation~\ref{eq:wc_odd}, for N=32 ions (solid line) as a function of the initial inter-ion distance.}
 \label{fig:wc_vs_d}
\end{figure}

A fourth order polynomial fit was found to correctly describe the evolution of $h$ after the transition has happened. For example, a fit on the slowest quench studied $\frac{\Delta \omega}{\tau_Q}=10^{6}[Hz/s]$, see fig~\ref{fig:N32_transition}c, leads to critical frequency of $\omega_c^{MD}/ 2\pi = 326.96kHz$, obtained by finding the roots of the polynomial. The same numerical value is obtained if the data from the quench $\frac{\Delta \omega}{\tau_Q}=10^{7}[Hz/s]$ is used instead. In order to save computational time, a quench of $\frac{\Delta \omega}{\tau_Q}=10^{7}[Hz/s]$ has been used to perform MD simulations for different initial values of inter-ion distance, using $N=32$ and $T=1nK$, leading to figure~\ref{fig:wc_vs_d}. The solid line of figure~\ref{fig:wc_vs_d} corresponds to eq~\ref{eq:wc_odd}, while the circles correspond to the critical frequencies deduced from the numerical simulations. The agreement between them is excellent.

\begin{figure}
 \centering
 \includegraphics[width=1.00\textwidth]{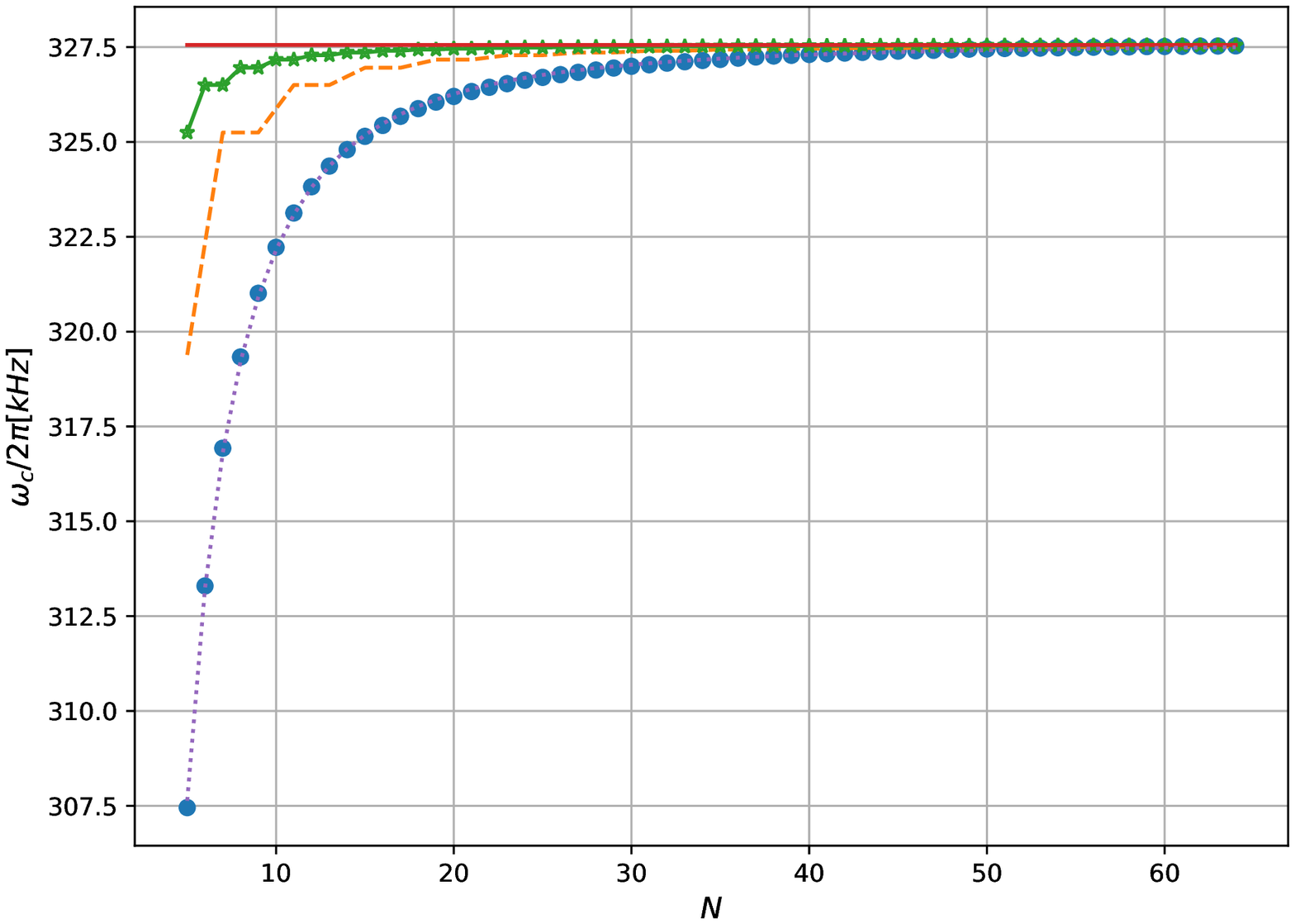}
 \caption{Comparison between the critical frequency obtained from MD simulations (full circles) and the analytical values given by equation~\ref{eq:wc_official_b} (red solid line), equation~\ref{eq:wc_official_a} (green stars) and equation~\ref{eq:wc_odd} (orange dashed line). The purple dotted line corresponds to a fit to the MD simulations results.}
 \label{fig:wc_vs_N}
\end{figure}

If now the inter-ion distance is fixed to $d=10\mu m$ and the number of ions is modified from $5$ to $64$, we obtain figure~\ref{fig:wc_vs_N}. In this figure the $\omega_c(d)$ from eq~\ref{eq:wc_official}, the computed values for $\omega_c(N,d)$ from eq~\ref{eq:wc_odd} and the results of the MD simulations are compared.

Equation~\ref{eq:wc_official_a} fails to reproduce the numerical results as expected. Although closer, equation~\ref{eq:wc_odd} also fails in predicting the right critical frequencies, particularly for smaller ion numbers. The reason becomes clear by comparing the structure assumed for the force calculation (full circles) and the structure at the end of the quenches (empty circles), as shown in figure~\ref{fig:ideal_zig_zag2}. The configuration used for the derivation of equation~\ref{eq:wc_odd} is an oversimplification and therefore the total transverse force experienced by the central ion has extra terms which were not accounted for, leading to a lower value of the critical frequency. Also, the numerical results show a continuous increase of $\omega_c$ with the number of ions, in disagreement with eqn.~\ref{eq:wc_odd} and eqn.~\ref{eq:wc_official_a} that have a non continuous increase. Nevertheless, eq~\ref{eq:wc_odd} gives a hint on the nature of the $N$ dependence. The following function has been use to fit the MD results:
\begin{equation}
\omega^2_{fit} = \frac{Q^2 k_c}{m d^3} a_1 \sum_{j=1}^{N}{\frac{1}{(j + a_2)^{a_3}}}
\end{equation}
leading to $a_1=(6.8\pm0.3)\cdot 10^{-5}$, $a_2=1.90\pm0.03$ and $a_3=3.22\pm0.02$. The fitted function shows an excellent agreement with MD simulations results.

\section{Defect generation during fast quenching in Homogeneous Chains}
The focus is shifted now to the generation of defects during a non-adiabatic but linear crossing of the transition. As discussed in the introduction, different causally disconnected regions of the chain can make different choices of the two possible configurations at the new phase. Fig.~\ref{fig:example_transition}c is an example of zig-zag, but the symmetric (respect $x=0$) configuration could equally have happened. At the boundary of two different topologies, a defect necessarily arises.

\begin{figure}
 \centering
 \includegraphics[width=1.00\textwidth]{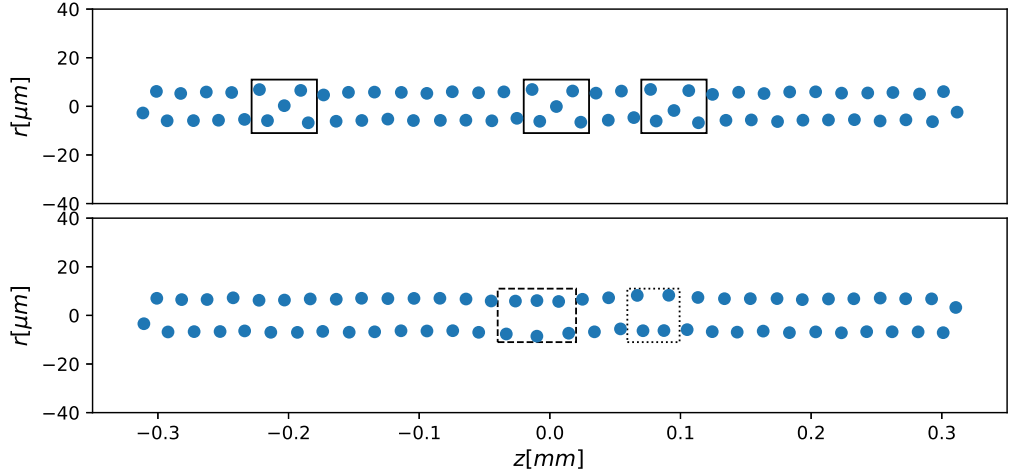}
 \caption{Random examples of final ion distribution after a quench. Several defects are indicated through boxes. Solid line box: \textit{intermediate} defect; dotted line box: \textit{odd} defect; dashed line box: \textit{extended} defect.}
 \label{fig:type_defects}
\end{figure}

Two examples of the final configurations for $N=64$ are shown in figure~\ref{fig:type_defects}. The defects can be classified in three categories~\cite{partner_dynamics_2013}: \textit{intermediate} defect, with one ion at the trap axis ($|x|\approx0$); \textit{odd} defect, where two consecutive ions positioned beside each other and \textit{extended} defects, where two ions have nearly the same axial position.

An algorithm that detects the number of defects by counting how many consecutive pairs have the same $x$ sign is used~\cite{algo_jofre}. Such an approach detects all three types of defects simultaneously. Its implementation is efficient and universal as it does not need manual input or threshold.

The results shown in figure~\ref{fig:N32_transition} indicates that the exact moment at which the phase transition takes place, and therefore the creation of eventual defects, depends on the quench rate used. Although the role of the temperature was not systematically explored, it is highly probable that it also plays a role to some extent on the exact frequency at which the transition takes place. The phase transition occurs after the adiabatic critical frequency has been crossed. It can even take place after the quench has finished. It implies that computing the number of defects at the end of the quench will lead to incorrect results. Therefore, in order to properly compare between different quench rates, we monitor the mean absolute transverse displacement, defined as $\langle x \rangle = \sum|x_i| / N$, and record the ion positions for post analysis when $\langle x \rangle$ reaches a particular fraction, $\epsilon$, of the adiabatic one, $\langle x \rangle = \epsilon \langle x\rangle_{adiabatic}$.

Another important aspect to take into account is defect loss which, unlike other previous works, it is clearly observable in the present study. Indeed, there are two mechanisms by which the number of defects can decrease during and after a quench: annihilation of the defect pairs and losses through the edges of the system. In order to gain insight on the evolution of the number of defects during a single quench, the average defect number was measured for $\epsilon=0.25$ to $\epsilon=0.85$ with steps of $d\epsilon=0.05$. For the smaller values of $\tau_Q$ studied, the quenches are too fast for the ions to reach the required values of $\epsilon$ during the quench duration itself. In those situations, the MD simulation is continued with $\omega = \omega_b$ until $\epsilon = 0.85$ is reached.

For the simulation results discussed henceforth, we have used $^{138}$Ba$^+$ ions, using a thermal bath of $T=0.5~$mK and a friction coefficient of $\Gamma=1.5\cdot10^{-20}~$ kg/s. The quench starts at $\omega_a / 2 \pi = 500$kHz and finishes at $\omega_b / 2 \pi = 140~$kHz. The values of the quench duration, $\tau_Q$ have been explored over a large enough range. Four different ion numbers have been studied: $N=30$, $N=64$, $N=128$, $N=256$. In all the cases, the ions are initialised with an ion-ion distance of $d= (10 + rand) \mu$m, where $rand$ is a random number in the range $\pm1$ and evolved for $2~$ms before the quench starts.

\begin{figure}
 \centering
 \includegraphics[width=1.00\textwidth]{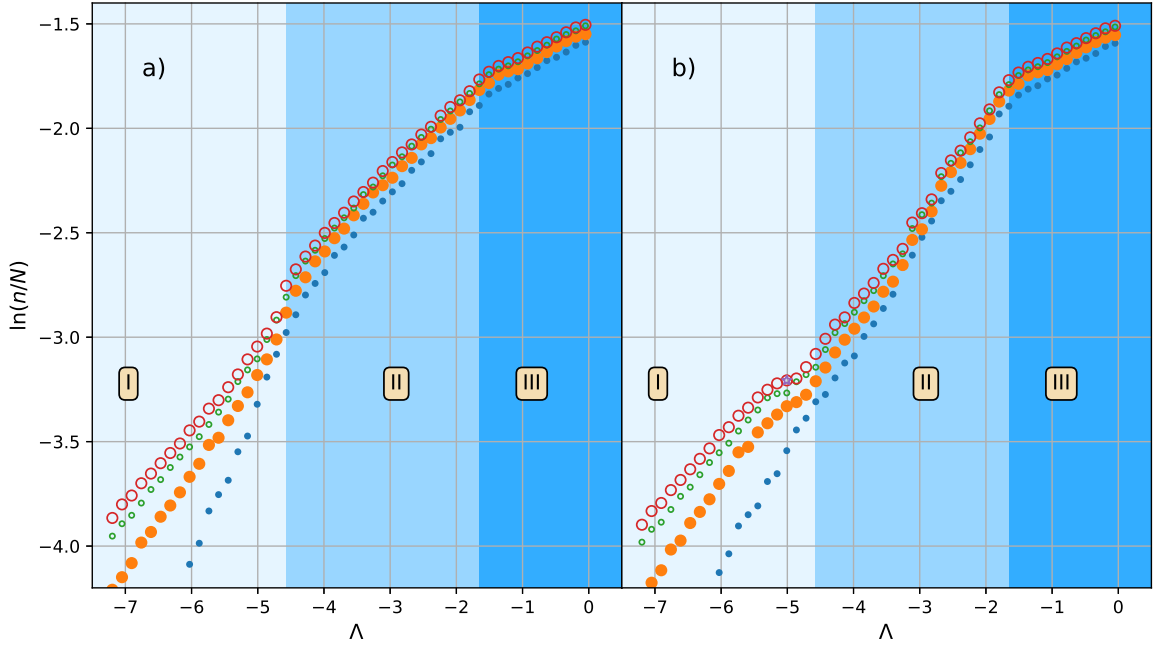}
 \caption{Number of defects versus the dimensionless quench rate (see main text). a): $\langle r \rangle = 0.25 \langle r_{a}\rangle$; b): $\langle r \rangle = 0.85 \langle r_{a}\rangle$. Small full circles: $N=30$; Large full circles: $N=64$; Small empty circles: $N=128$; Large empty circles: $N=256$. The standard deviation of each point is smaller than the symbols used.}
 \label{fig:CompareN_first_last}
\end{figure}

The evolution of the logarithm of the averaged normalized number of defects, $\ln(n/N)$, versus the logarithm of the dimensionless quench rate, $\Lambda = -\ln(\tau_Q \omega_0)$, with $\omega_0^2 = \frac{Q^2 k_C}{m d^3}$~\cite{nigmatullin_formation_2016}, is shown in figure~\ref{fig:CompareN_first_last} for $\epsilon=0.25$ and $\epsilon=0.85$. The results for the different ion numbers, $N$, are shown. Each point on figure~\ref{fig:CompareN_first_last} corresponds to the average of $2000$ independent simulations. The standard deviations on the normalized number of defects, $n/N$, ranges from $10^{-4}$ and $10^{-3}$. Those values are much smaller than the symbols used on figure~\ref{fig:CompareN_first_last}. 

Three regions can be identified in figure~\ref{fig:CompareN_first_last}. Region III where fast quench occurs, does not show a significant difference on the number of defects between $\epsilon=0.25$ and $\epsilon=0.85$. The values for  $\ln(n/N)$, and therefore the slope, are similar among the different studied ion numbers. This is in contrast with the slow quench regime, region I, where although there is no significant defect loss between $\epsilon=0.25$ and $\epsilon=0.85$, there is a clear dependence on the ion number.

The absence of losses in region I can be explained by a very slow of dynamics of the defects generated, slower than the time difference between $\epsilon=0.25$ and $\epsilon=0.85$ or that any defect dynamics, and therefore any loses, have already taken place at $\epsilon=0.25$ due that the slow quench regime of this region. For fast quenches, region III, the absence of losses could be explained by the fact that the times involved are too short for any dynamics of the defects to take place.

Region II presents a clear evolution of the number of defects with the value of $\epsilon$: a significant portion of the defects present at $\epsilon=0.25$ have been lost when $\epsilon=0.85$ has been reached. Moreover, while the power-law is similar for region II at $\epsilon=0.25$ across the different values of the ion number $N$ studied, see figure~\ref{fig:CompareN_first_last}a, the slope increases at $\epsilon=0.85$ in region II as the ion chain size decreases, see figure~\ref{fig:CompareN_first_last}b.

This behaviour can be interpreted using the work of Partner et. al~\cite{partner_dynamics_2013} on the dynamical behaviour of the defects created in the context of single defects in laser cooled ions on "standard" ion traps, where the harmonic potential along the axis leads to the creation of defects in the middle of the chain. Through the calculation of the Peierls-Nabarro (PN) potential, Partner et. al~\cite{partner_dynamics_2013} was able to explain qualitatively the dynamics of such defects. The PN potential corresponds to the potential seen by the defect along the chain as a function of the defect center (see~\cite{partner_dynamics_2013} for a definition of defect center and how to compute the PN potential). The PN potential rises toward the edges for the \textit{extended} defects while decreases for the other two types, \textit{odd} and \textit{intermediate} defects. Moreover, Partner et. al~\cite{partner_dynamics_2013} also showed how defect dynamics imply the transformation among the different types of defects and, as a consequence, the defect sees alternatively a confining and de-confining PN potential. While the above results were obtained in an ion chain trapped using an axial potential and with a non-uniform ion distribution, the PN potential on an homogenous chain, although flatter at the center, should be qualitatively similar.

Full ion trajectories were recorded for some single quench rates. Their analysis showed axial displacements of the defects, sometimes leading to losses through the edges while other times the defect oscillated along the chain without being lost. Such observations have no statistical meaning, but give us some hint of the type of dynamics that can be expected and they are in accordance to what was reported in~\cite{partner_dynamics_2013}. It also provides insight on how edge plays a role in defect dynamics.

Therefore, the difference in the power laws observed in region II when $\epsilon=0.85$, can be explained by the fact that, as we are working in an homogeneous system, the probability of defect creation is uniform along the chain and therefore the probability that a defect is created at the edge is higher for smaller systems than for larger systems. Once the defects are created, the ones closer to the edge are more easily lost if their corresponding PN potential is a de-confining one.

The same reason can be invoked for the \textit{knee} observed between region II and region I observed for $\epsilon=0.85$ and not for $\epsilon=0.25$. The strong \textit{knee} observed for $N=256$ at $\epsilon=0.85$, coupled with its absence at $\epsilon=0.25$ can be explained due to a decrease of the absolute number of defects between both values of $\epsilon$, implying a decrease of defect density in the chain, which in turns means that the probability of two defects annihilating each other during a quench should decrease. Such interpretation also explains the decrease of the \textit{knee} with the decrease of the number of ions: the relative annihilation contribution to the defect losses during a quench is necessarily lower for smaller chains.

\begin{figure}
 \centering
 \includegraphics[width=1.00\textwidth]{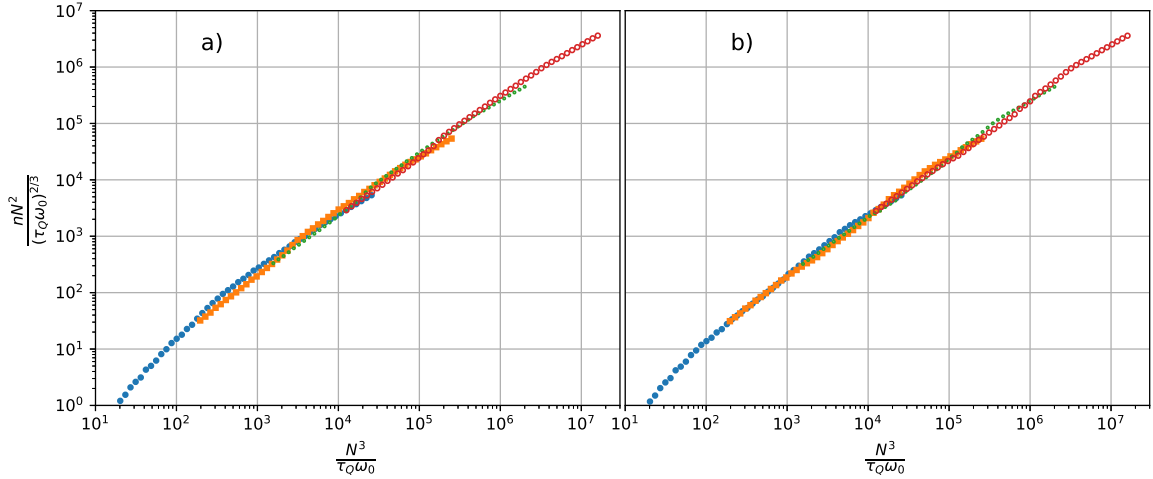}
 \caption{Plot of $\frac{n N^2}{(\tau_Q \omega_0)^{2/3}}$ vs $\frac{N^3}{\tau_Q \omega_0}$. a) $\epsilon=0.25$, b) $\epsilon=0.85$. Blue circles: $N=30$; Orange squares: $N=64$; Green circles: $N=128$; Empty Red circles: $N=256$.}
 \label{fig:rescaling_Nigmatullin}
\end{figure}

Finally, it should be noted that by normalizing the number of defects with respect to the total number of ions, $n/N$, clearly converges as the ion number increases. The numerical values for $N=128$ and $N=256$ are indeed very close, specially for region II and III when $\epsilon=0.25$, indicating the approximation of an infinite system should be valid at this regions and for such value of $\epsilon$. The finite size universality for an homogeneous system can be recovered by using a different scaling if $\frac{n N^2}{(\tau_Q \omega_0)^{2/3}}$ vs $\frac{N^3}{\tau_Q \omega_0}$ is plotted, as it was shown by Nigmatullin et al.~\cite{nigmatullin_formation_2016}. Our results, plotted using such scaling, are shown in figure~\ref{fig:rescaling_Nigmatullin}. The different ion number results collapse, as an overall trend, into a single line thus confirming the results from Nigmatullin et al. However, a closer look clearly indicates some deviations, coming from slow quenches set of results. The differences from our results and the ones reported in~\cite{nigmatullin_formation_2016} is that in Nigmatullin et al. periodic boundary conditions are imposed. These boundary conditions forbids any type of defect loss as the PN potential will not decrease at the edges, since there exists no edges, and therefore, no defect dynamics are expected. 

\section{Homogeneous KZM}
The present work differs from the standard HKZM in two main aspects. First, the scaling law derived from HKZM theory for the number of defects generated does not take into account possible defect losses. Secondly, the presence of laser cooling, modelised here as a damping term, was not present in the original KZM theory. The introduction of a damping term has been studied within the Ginzburg-Landau description by several authors, for example Chiara et al.~\cite{chiara_spontaneous_2010}. Two specific limits have been analysed on the context of infinite homogeneous ion chains~\cite{chiara_spontaneous_2010}: the over-damped and the under-damped, leading to $n \propto \left(\tau_Q \omega_0\right)^{-1/4}$ and $n \propto \left(\tau_Q \omega_0\right)^{-1/3}$, respectively.

\begin{figure}
 \centering
 \includegraphics[width=1.0\textwidth]{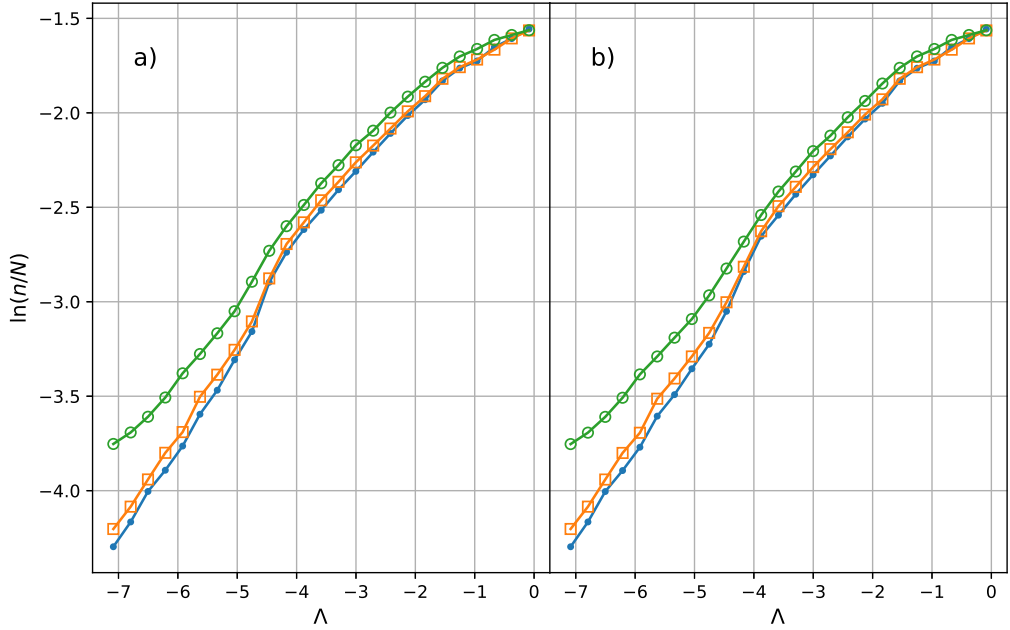}
 \caption{Number of defects versus $\Lambda$ and $\epsilon$ for several values of the friction coefficient, $\Gamma$, for $N=64$ and two values of $\epsilon$: a) $\epsilon=0.25$ and b) $\epsilon=0.85$. Blue Small dots: $\Gamma=1.5\cdot10^{-21}kg/s$; Orange Empty Squares: $\Gamma=1.5\cdot10^{-20}kg/s$; Green Empty Circles: $\Gamma=1.5\cdot10^{-19}kg/s$. }
 \label{fig:comparison_5mK_all_eta}
\end{figure}

It is this second scenario, the under-damped limit, that leads to the same scaling as the standard HKZM. In Nigmatullin et al~\cite{nigmatullin_formation_2016}, it is argued that in the under-damped limit, a small change on the damping term should not significantly affect the results. For this reason, we have performed MD simulations for three different values of $\Gamma$: $1.5\cdot10^{-21} kg/s$, $1.5\cdot10^{-20} kg/s$ and $1.5\cdot10^{-19} kg/s$. For this comparison only $N=64$ have been used. The rest of the parameters and procedure remain the same as before. The results, figure~\ref{fig:comparison_5mK_all_eta}, show how the results corresponding to $\Gamma=1.5\cdot10^{-21} kg/s$ and $1.5\cdot10^{-20} kg/s$ are essentially the same. Therefore, our simulation parameters over the quench rate range explored can be safely assumed to belong to the under-damped limit.

\begin{figure}[t]
 \centering
 \includegraphics[width=1.00\textwidth]{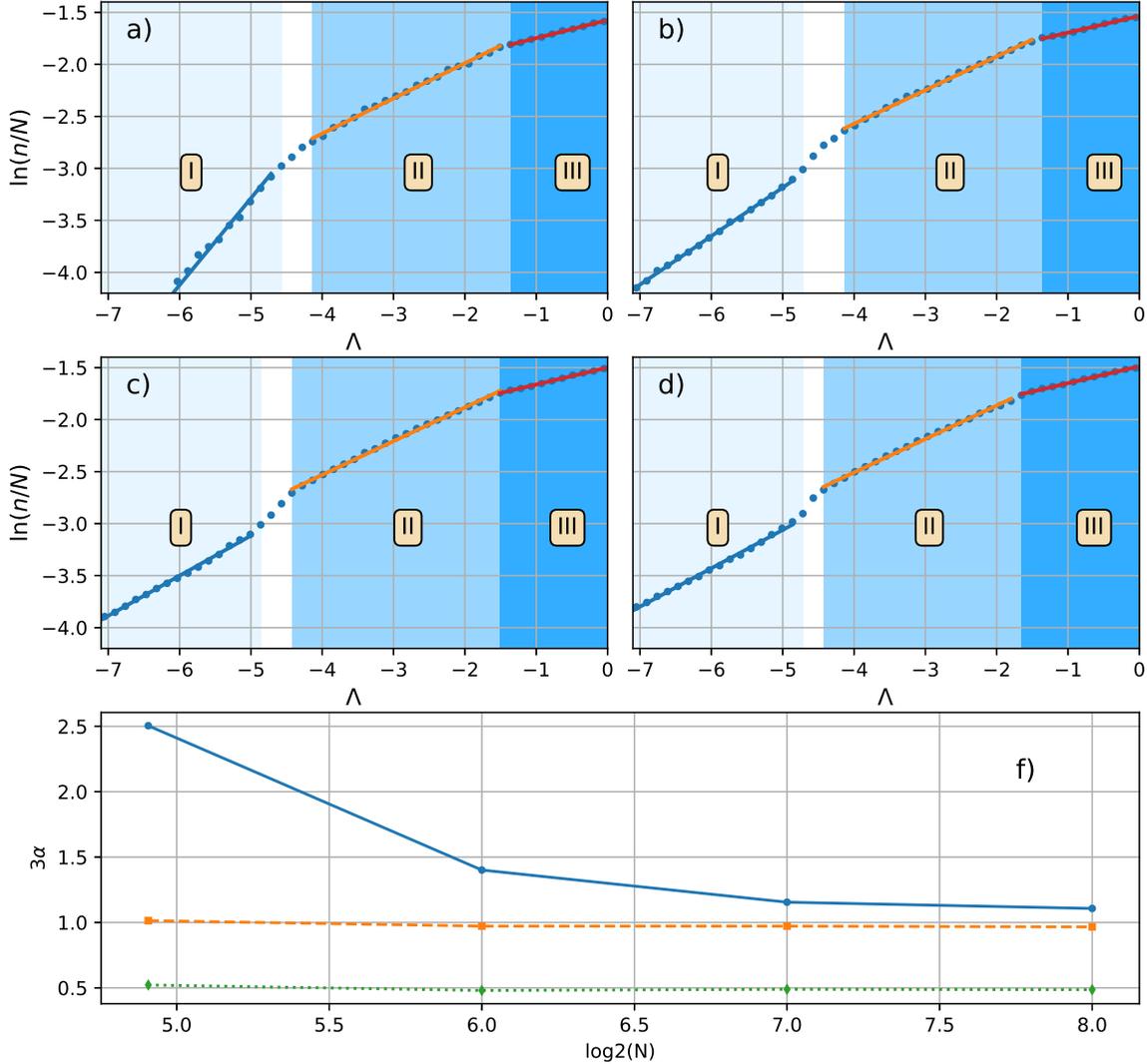}
 \caption{Number of defects versus the adimensional quench rate for different values of N: a)N=30; b)N=64; c)N=128; d)N=256. The power low fit to each region is shown. f) The values of the fit (multiplied by 3) are shown in function of the number of ions. Circles: region I; Squares: region II; Diamonds: region III. Notice that the error bars associated with the fits are much smaller than the symbols.}
 \label{fig:fits}
\end{figure}

In order to minimise the effects of defect loss, the curve corresponding to $\epsilon=0.25$, figure~\ref{fig:CompareN_first_last}a, has been used to obtain the coefficient of the power law for the different regions and different number of ions, $n \propto \left(\tau_Q \omega_0\right)^{-\alpha}$, see figure~\ref{fig:fits}.

Figure~\ref{fig:fits}c show how region II agrees with the expected value corresponding to the HKZM, $\alpha = 1/3$~\cite{del_campo_universality_2014}: $\alpha_{II}[N=30] = 0.338\pm0.005$, $\alpha_{II}[N=64] = 0.324\pm0.004$, $\alpha_{II}[N=128] = 0.324\pm0.005$ and $\alpha_{II}[N=256] = 0.322\pm0.004$. While the agreement with the expected value is relatively good, the values are consistently but only slightly lower. The reason is probably that the results are already affected by defect losses at $\epsilon=0.25$. In addition, it should be noticed that these values are sensitive to the exact interval used and they can change significantly.

Interestingly, a new scaling law seems to appear at region III $\alpha_{III} \approx 1/6$. The values are $\alpha_{III}[N=30] = 0.174 \pm 0.003$, $\alpha_{III}[N=64] = 0.160 \pm 0.005$; $\alpha_{III}[N=128] = 0.163 \pm 0.003$ and $\alpha_{III}[N=256] = 0.162 \pm 0.003$. The origin of this new scaling, lower to than the standard HKZM, is unknown although some possible explanations are hinted in the next section.

Finally, three different regions were already observed in the case of the IKZM, see figure~2 of Pyka et al.~\cite{pyka_topological_2013}, corresponding to 1/3, 4/3 and 8/3 for (qualitatively speaking) fast, intermediate and slow quenches respectively. The change from 1/3 to 4/3 appears from the restriction of the region where defects can appear, due to the highest density of ions at the ion chain centre when an harmonic potential is used to confine the ions~\cite{del_campo_structural_2010}. In the present case, the homogeneity of the initial ion chain assures that the defect formation is not restricted spatially, independently of the quench rate used.

For slow quenches, the correlation length, $\xi$, defined in the present context as the distance between two consecutive defects, becomes comparable with the size of the system and the density of defects equals the probability of a single defect, leading to a doubling of the previous power law to 8/3~\cite{del_campo_structural_2010}. The same argument could be invoked here as a factor 2.5 appears for the smaller chain studied, $N=30$. As the length of the ion chain increases, the power law of this slow quench region approaches 1/3 of the HKZM. This is consistent with the evolution of the correlation length respect the system size, $\frac{\xi}{L} = \frac{\xi}{d(N-1)}$ (see next section), where the correlation length decreases as the ion number increases for the slowest quenches studied.

\section{Correlation length}
Deviations from the standard HKZM are expected to happen when the correlation length is comparable to the length scale of the system, in this case the ion-ion distance, as the low-energy Ginzburg-Landau theory is no longer valid. At the other extrema, for very slow quenches, the correlation length could be of similar order as our finite size system, the ion chain length. 

\begin{figure}[t]
 \centering
 \includegraphics[width=1.00\textwidth]{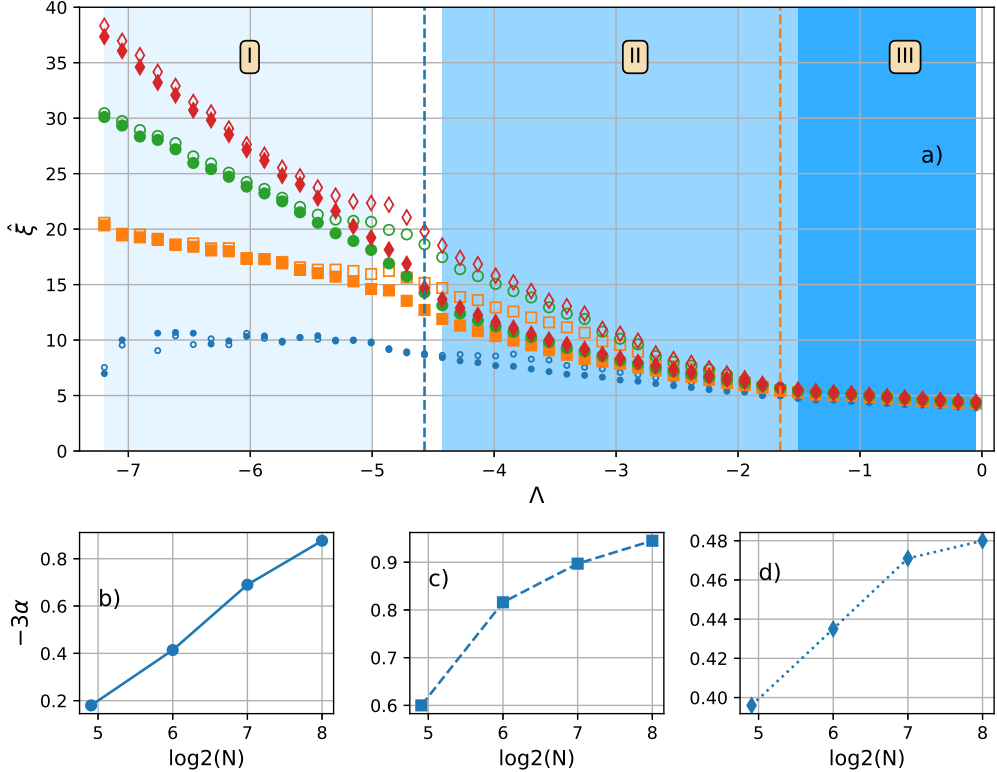}
 \caption{a) Normalised correlation length vs normalised quench rate for the different ion numbers studied at two different values of $\epsilon$: full symbols,$\epsilon=0.25$; empty symbols $\epsilon=0.85$. Blue small circles: $N=30$; orange squares: $N=64$; green large circles: $N=128$; red diamonds: $N=256$. The two vertical lines indicate the boundaries between the regions observed in figure~\ref{fig:CompareN_first_last}. The b), c) and d) figures correspond to the coefficient of the power law fits for region I, region II and region III respectively.}
 \label{fig:corr_length}
\end{figure}

Therefore, the evolution of the correlation length has also been extracted from the MD simulations. The mean values of the correlation length for each quench time (corresponding to $\epsilon=0.25$ for full symbols and to $\epsilon=0.85$ for empty symbols), normalised by the ion-ion distance, $\hat{\xi} = \xi / d$ are shown in figure~\ref{fig:corr_length}a where the standard deviation is smaller than the size of the symbols used. The distances from the first and last defect to the edges have been excluded. This new figure shows how the average correlation length increases with the parameter $\epsilon$, and therefore, with time.

Once more, several regions with different behaviours are observed. The regions do not share the same boundaries for the different ion numbers and therefore, the regions marked with different tones of blue indicate the data sets used for the power law fits. The two vertical lines indicate the boundaries between the regions observed in figure~\ref{fig:CompareN_first_last}.

When comparing with figure~\ref{fig:CompareN_first_last}, we remark that the regions where the $\epsilon=0.25$ and $\epsilon=0.85$ differs, are not the same for the $\Lambda$ and $\hat{\xi}$. This is specially marked for slow quenches. For example, for $N=30$ and $\Lambda=-5$, where the averaged correlation length does not change as $\epsilon$ increases but the number of defects does decrease as $\epsilon$ increases. If there was annihilation of pair of defects, the correlation length should increase, which is not the case. With the current information at hand, a constant correlation length coupled with a loss of defects can only be understood as a collective behaviour happening at the edges. Further theory as well as simulation work is needed to elucidate the observed behaviour which is beyond the scope of this work.

Finally, it seems as the adiabatic correlation length is only achieved for $N=30$, where the correlation length shows a plateau at slow quenches corresponding roughly to $L/3$. If the $L/3$ should also be applied to the other studied cases, a plateau should be observed at: $\approx 21$ for $N=64$, $\approx 43$ for $N=128$ and $\approx 85$ for $N=256$. We see that $N=64$ was just reaching $\hat{\xi} \approx 21$ for the slowest of the simulated quench rates, while the higher values of $N$, were still far from it. The theoretical value obtained using using Ginzburg-Landau theory ~\cite{puebla_fokker-planck_2017}: $\frac{1}{2\sqrt{6}}\approx 0.204$, which is lower than our measured value of $1/3$.

The scaling law of $1/6$ observed in region III of figure~\ref{fig:CompareN_first_last} corresponds approximately to $\hat{\xi} < 5.5$. Such low value of the $\hat{\xi}$ could explain the change on the scaling as the model used for the derivation of the HKZM loses its validity~\cite{nigmatullin_formation_2016}.

A power law fit to the correlation length for the different regions defined by the different blue regions in figure~\ref{fig:corr_length}a has been done. The HKZM predicts an scaling of $\hat{\xi} \sim \left(\tau_Q \omega_0\right)^{1/3}$. Region II, represented as squares in figure~\ref{fig:corr_length}c, seem to approach such a value as the ion chain increases. The correlation length appears to be more sensitive to finite size effects than the number of defects as the value has not yet converged to the infinite case, $1/3$ when $N=256$. The power law in region III, figure~\ref{fig:corr_length}c, seem to converge again to the same scaling of $1/6$, already observed in the defect density in this region.

\section{Conclusion}
We investigated the defect dynamics during a 1D to 2D phase transition of an uniform ion chain confined in a segmented linear ion trap using MD simulations. This allowed us for the first time to study a homogeneous but finite size system without the need of periodic boundary conditions, and therefore introducing the possibility of defect dynamics in presence of defect loses.
 
A new analytic expression for the critical frequency at which a uniform linear chain undergoes a structural phase transition to a zig-zag has been derived. Using MD simulations, the critical frequency has been obtained for a range of different ion numbers and ion-ion distances. The numerical results showed a better agreement with the new expression but the deviation for small ion numbers is still significant. However, a fit to a function inspired by this new expression, leads to an excellent agreement. The MD simulations showed that the rate at which the control parameter is changed, modifies significantly the critical frequency at which the transition takes place, motivating the introduction of the $\epsilon$ parameter.

The role of the $\epsilon$ parameter has proven critical in order to understand defect loses dynamics and to identify the quench rate parameter range over which the HKZM scaling can be obtained. Furthermore, it allowed us to identify three different regimes regarding the defect generation scaling and the defect losses for a given quench rate. In the region that better agrees with the expected HKZM scaling, defect loss have been shown to depend on the system size, given here by the ion number $N$.

We further investigated the dependence of the correlation length evolution with the quench rate. This combined with the loss dynamics for the same quench rate indicates the appearance of new collective defect behaviour at edges. Finally, we have shown that the HKZM regime emerges with a relative low ion number, $N=30$. Such ion number value is experimentally accessible and it could open the door to an HKZM experimental verification on a system with an extreme controllability and with a high degree of repeatability, essential for the building up of the needed statistics of such types of experiments.\\


\begin{thebibliography}{10}
\expandafter\ifx\csname url\endcsname\relax
  \def\url#1{{\tt #1}}\fi
\expandafter\ifx\csname urlprefix\endcsname\relax\def\urlprefix{URL }\fi
\providecommand{\eprint}[2][]{\url{#2}}

\bibitem{zurek_cosmological_1985}
Zurek W~H 1985 {\em Nature\/} {\bf 317} 505--508 ISSN 1476-4687 number: 6037
  Publisher: Nature Publishing Group
  \urlprefix\url{https://www.nature.com/articles/317505a0}

\bibitem{kibble_topology_1976}
Kibble T~W~B 1976 {\em Journal of Physics A: Mathematical and General\/} {\bf
  9} 1387--1398 ISSN 0305-4470 publisher: IOP Publishing
  \urlprefix\url{https://iopscience.iop.org/article/10.1088/0305-4470/9/8/029}
  
\bibitem{del_campo_universality_2014}
del Campo A and Zurek W~H 2014 {\em International Journal of Modern Physics
  A\/} {\bf 29} 1430018 ISSN 0217-751X, 1793-656X arXiv: 1310.1600
  \urlprefix\url{http://arxiv.org/abs/1310.1600}

\bibitem{pyka_topological_2013}
Pyka K, Keller J, Partner H~L, Nigmatullin R, Burgermeister T, Meier D~M,
  Kuhlmann K, Retzker A, Plenio M~B, Zurek W~H, del Campo A and Mehlstäubler
  T~E 2013 {\em Nature Communications\/} {\bf 4} 2291 ISSN 2041-1723
  \urlprefix\url{http://www.nature.com/articles/ncomms3291}

\bibitem{ulm_observation_2013}
Ulm S, Roßnagel J, Jacob G, Degünther C, Dawkins S~T, Poschinger U~G,
  Nigmatullin R, Retzker A, Plenio M~B, Schmidt-Kaler F and Singer K 2013 {\em
  Nature Communications\/} {\bf 4} 2290 ISSN 2041-1723
  \urlprefix\url{http://www.nature.com/articles/ncomms3290}

\bibitem{kamsap_experimental_2017}
Kamsap M~R, Champenois C, Pedregosa-Gutierrez J, Mahler S, Houssin M and Knoop
  M 2017 {\em Physical Review A\/} {\bf 95} 013413 ISSN 2469-9926, 2469-9934
  \urlprefix\url{https://link.aps.org/doi/10.1103/PhysRevA.95.013413}

\bibitem{del_campo_structural_2010}
del Campo A, De~Chiara G, Morigi G, Plenio M~B and Retzker A 2010 {\em Physical
  Review Letters\/} {\bf 105} 075701 ISSN 0031-9007, 1079-7114
  \urlprefix\url{https://link.aps.org/doi/10.1103/PhysRevLett.105.075701}

\bibitem{champenois_ion_2010}
Champenois C, Marciante M, Pedregosa-Gutierrez J, Houssin M, Knoop M and Kajita
  M 2010 {\em Physical Review A\/} {\bf 81} 043410 ISSN 1050-2947, 1094-1622
  \urlprefix\url{https://link.aps.org/doi/10.1103/PhysRevA.81.043410}

\bibitem{nigmatullin_formation_2016}
Nigmatullin R, del Campo A, De~Chiara G, Morigi G, Plenio M~B and Retzker A
  2016 {\em Physical Review B\/} {\bf 93} 014106 ISSN 2469-9950, 2469-9969
  \urlprefix\url{https://link.aps.org/doi/10.1103/PhysRevB.93.014106}

\bibitem{pedregosa-gutierrez_symmetry_2018}
Pedregosa-Gutierrez J, Champenois C, Kamsap M~R, Hagel G, Houssin M and Knoop M
  2018 {\em Journal of Modern Optics\/} {\bf 65} 529--537 ISSN 0950-0340,
  1362-3044
  \urlprefix\url{https://www.tandfonline.com/doi/full/10.1080/09500340.2017.1408866}

\bibitem{pedregosa-gutierrez_correcting_2018}
Pedregosa-Gutierrez J, Champenois C, Houssin M, Kamsap M~R and Knoop M 2018
  {\em Review of Scientific Instruments\/} {\bf 89} 123101 ISSN 0034-6748
  publisher: American Institute of Physics
  \urlprefix\url{https://aip.scitation.org/doi/10.1063/1.5075496}

\bibitem{li_realization_2017}
Li H~K, Urban E, Noel C, Chuang A, Xia Y, Ransford A, Hemmerling B, Wang Y, Li
  T, Häffner H and Zhang X 2017 {\em Physical Review Letters\/} {\bf 118}
  053001 ISSN 0031-9007, 1079-7114
  \urlprefix\url{https://link.aps.org/doi/10.1103/PhysRevLett.118.053001}

\bibitem{lin_large-scale_2009}
Lin G~D, Zhu S~L, Islam R, Kim K, Chang M~S, Korenblit S, Monroe C and Duan L~M
  2009 {\em EPL (Europhysics Letters)\/} {\bf 86} 60004 ISSN 0295-5075,
  1286-4854
  \urlprefix\url{https://iopscience.iop.org/article/10.1209/0295-5075/86/60004}

\bibitem{raino_superfluorescence_2018}
Rainò G, Becker M~A, Bodnarchuk M~I, Mahrt R~F, Kovalenko M~V and Stöferle T
  2018 {\em Nature\/} {\bf 563} 671--675 ISSN 1476-4687 number: 7733 Publisher:
  Nature Publishing Group
  \urlprefix\url{https://www.nature.com/articles/s41586-018-0683-0}

\bibitem{li_structural_2020}
Li S, Duan S, Zha Z, Pan J, Sun L, Liu M, Deng K, Xu X and Qiu X 2020 {\em ACS
  Nano\/} ISSN 1936-0851 publisher: American Chemical Society
  \urlprefix\url{https://doi.org/10.1021/acsnano.0c02995}

\bibitem{puebla_fokker-planck_2017}
Puebla R, Nigmatullin R, Mehlstäubler T~E and Plenio M~B 2017 {\em Physical
  Review B\/} {\bf 95} 134104 ISSN 2469-9950, 2469-9969
  \urlprefix\url{http://link.aps.org/doi/10.1103/PhysRevB.95.134104}

\bibitem{johanning_isospaced_2016}
Johanning M 2016 {\em Applied Physics B\/} {\bf 122} 71 ISSN 0946-2171,
  1432-0649 \urlprefix\url{http://link.springer.com/10.1007/s00340-016-6340-0}

\bibitem{leung_entangling_2018}
Leung P~H and Brown K~R 2018 {\em Physical Review A\/} {\bf 98} 032318 ISSN
  2469-9926, 2469-9934
  \urlprefix\url{https://link.aps.org/doi/10.1103/PhysRevA.98.032318}

\bibitem{xie_creating_2017}
Xie Y, Zhang X, Ou B, Chen T, Zhang J, Wu C, Wu W and Chen P 2017 {\em Physical
  Review A\/} {\bf 95} 032341 ISSN 2469-9926, 2469-9934
  \urlprefix\url{http://link.aps.org/doi/10.1103/PhysRevA.95.032341}

\bibitem{skeel_impulse_2002}
Skeel R~D and Izaguirre J~A 2002 {\em Molecular Physics\/} {\bf 100} 3885--3891
  ISSN 0026-8976, 1362-3028
  \urlprefix\url{http://www.tandfonline.com/doi/abs/10.1080/0026897021000018321}

\bibitem{fishman_structural_2008}
Fishman S, De~Chiara G, Calarco T and Morigi G 2008 {\em Physical Review B\/}
  {\bf 77} 064111 ISSN 1098-0121, 1550-235X
  \urlprefix\url{https://link.aps.org/doi/10.1103/PhysRevB.77.064111}

\bibitem{shimizu_dynamics_2018}
Shimizu K, Kuno Y, Hirano T and Ichinose I 2018 {\em Physical Review A\/} {\bf
  97} 033626 ISSN 2469-9926, 2469-9934
  \urlprefix\url{https://link.aps.org/doi/10.1103/PhysRevA.97.033626}

\bibitem{dutta_nonequilibrium_2012}
Dutta T, Mukherjee M and Sengupta K 2012 {\em Physical Review A\/} {\bf 85}
  063401 ISSN 1050-2947, 1094-1622
  \urlprefix\url{https://link.aps.org/doi/10.1103/PhysRevA.85.063401}

\bibitem{partner_dynamics_2013}
Partner H~L, Nigmatullin R, Burgermeister T, Pyka K, Keller J, Retzker A,
  Plenio M~B and Mehlstäubler T~E 2013 {\em New Journal of Physics\/} {\bf 15}
  103013 ISSN 1367-2630
  \urlprefix\url{http://stacks.iop.org/1367-2630/15/i=10/a=103013?key=crossref.5fbb9cda352eac4328a040649440f888}

\bibitem{algo_jofre}
 {\em The algorithm can be written in a single line when using Python/Scipy: \\
  n = size(where(abs(diff(sign(x)))~\textless 1)).\/}

\bibitem{chiara_spontaneous_2010}
Chiara G~D, Campo A~d, Morigi G, Plenio M~B and Retzker A 2010 {\em New Journal
  of Physics\/} {\bf 12} 115003 ISSN 1367-2630
  \urlprefix\url{http://stacks.iop.org/1367-2630/12/i=11/a=115003?key=crossref.54fb0240f2226d035f2456b8f4a8b778}

\end{thebibliography}

\providecommand{\newblock}{}

\end{document}